\documentclass[twocolumn,twoside]{IEEEtran} 

\usepackage{multirow}
\usepackage{amsmath,amssymb,amsfonts}
\usepackage[final]{graphicx}
\usepackage{psfrag}
\usepackage[numbers,sort&compress]{natbib}

\usepackage{flushend}

\usepackage[boxed]{algorithm2e}

\newtheorem{corollary}{Corollary}

\newtheorem{lemma}{Lemma}
\newtheorem{definition}{Definition}
\usepackage{color, soul}

\def \E {\mathbb E}

\makeatletter
\def\blfootnote{\xdef\@thefnmark{}\@footnotetext}
\makeatother

\begin{document}
\title{Unveiling the Hyper-Rayleigh Regime of the Fluctuating Two-Ray Fading Model}
\hyphenation{hyper-geo-met-ric shad-ow-ing}

\author{\IEEEauthorblockN{Celia Garcia-Corrales, Unai Fernandez-Plazaola, Francisco J. Ca\~nete, Jos\'e F. Paris and F. Javier Lopez-Martinez}
\thanks{This work has been submitted to the IEEE for publication. Copyright may be transferred without notice, after which this version may no longer be accesible.}
\thanks{The authors are with Departmento de Ingenier\'ia de Comunicaciones, Universidad de M\'alaga - Campus de Excelencia Internacional Andaluc\'ia Tech., M\'alaga 29071, Spain (e-mail: {\rm \{celia,unai,francis,paris,fjlopezm\}@ic.uma.es}).}
\thanks{This work has been funded by the Spanish Government and the European Regional Development Fund (TEC2014-57901-R and TEC2017-87913-R), and also by the University of M\'alaga.}
}

\maketitle

\begin{abstract}
The recently proposed Fluctuating Two-Ray (FTR) model is gaining momentum as a reference fading model in scenarios where two dominant specular waves are present. Despite the numerous research works devoted to the performance analysis under FTR fading, little attention has been paid to effectively understanding the interplay between the fading model parameters and the fading severity. According to a new scale defined in this work, which measures the hyper-Rayleigh character of a fading channel in terms of the Amount of Fading, the outage probability and the average capacity, we see that the FTR fading model exhibits a full hyper-Rayleigh behavior. However, the Two-Wave with Diffuse Power fading model from which the former is derived has only strong hyper-Rayleigh behavior, which constitutes an interesting new insight. We also identify that the random fluctuations in the dominant specular waves are ultimately responsible for the full hyper-Rayleigh behavior of this class of fading channels.
\end{abstract}


\section{Introduction}


With the advent of the new century, the research in stochastic fading models has been revamped as a complement to the classical models like Rayleigh, Rice and Nakagami. A number of relevant and rather general fading distributions has been proposed \cite{Durgin2002,Abdi2003, Yacoub2007, Paris2014,Romero2017}, which have proven useful to accommodate to a wide set of propagation environments and have been supported by empirical evidences. 

Among these new channel models, the Fluctuating Two-Ray (FTR) fading model \cite{Romero2017} has become rather popular due to its versatility to represent propagation conditions on which not one, but two dominant waves appear, associated to specular multipath components. These two dominant waves are often referred to as line-of-sight (LoS) in a wide sense. Right after its inception, the research interest on the FTR fading model has been intense, providing further generalizations of it and characterizing the performance of wireless communication systems operating under FTR fading channels \cite{Zhang2018,Zhao2019,Zeng2018}.

The FTR model inherits the characteristics of the Two-Wave with Diffuse Power (TWDP) model \cite{Durgin2002} from which it originates. The TWDP model contemplates a Rayleigh-diffused component which is added to the two deterministic LoS ones. This provides additional degrees of freedom for the desired model to accurately represent the actual channel conditions compared to the Rician case (i.e. only one LoS component). The two dominant specular waves could be clearly discernible (depending on electromagnetic issues like antennas radiation pattern, signal bandwidth or carrier frequency) over other contributions from waves of smaller amplitudes due to multipath propagation in the environment that could be grouped into the diffuse term. The key innovation of FTR w.r.t. the TWDP model relies on the fact that the LoS components are allowed to fluctuate. This may be caused, for instance, by human body shadowing \cite{Cotton2011}, which occurs at a faster time-scale than the \emph{conventional} shadowing associated to the presence of buildings. Examples of suitable propagation scenarios for these models can be found in indoor mobile radio links, vehicle-to-vehicle communications, wireless sensor networks or mmWave communications \cite{Matolak2011,Frolik2007,Romero2017}. 

Rayleigh fading model is naturally derived from an scenario of narrow-band wireless transmission with a large number of uncorrelated components, due to multipath propagation, of similar amplitudes and without a dominant wave associated to a LoS path \cite{Simon2005} (a Non-LoS --NLoS-- condition). Hence, it is common to use it as a benchmark to compare other fading models, and it has been traditionally considered as a worst case situation. The term worse-than-Rayleigh has been coined to denote channel models with a more severe fading, which yields lower system performance than the Rayleigh counterpart \cite{Sen2006, Matolak2011}. This notion is commonly associated to scenarios in which several dominant waves (more than one, which corresponds to Rice fading) of similar amplitudes may cancel each other, i.e. like TWDP fading. Also, multipath propagation conditions in which the number of components is not large enough to apply the central limit theorem as in the Rayleigh model are expected to behave in this manner. The term hyper-Rayleigh is often employed as a synonym for this worse-than-Rayleigh behavior \cite{Frolik2007,Frolik2008}, 

The TWDP fading, and its special case Two-Wave, are widely used as examples of hyper-Rayleigh behavior in the literature \cite{Frolik2007,Frolik2008,Rao2015}. Because it is derived from the TWDP fading model, the FTR model is expected to exhibit also a worse-than-Rayleigh behavior, associated to the partial cancellation of the LoS components when they have similar magnitudes and the diffuse power is reduced (i.e. strong LoS case). Even though some attempts have been made to investigate the hyper-Rayleigh behavior in the literature from different perspectives, there is still no consensus in the community to quantify this effect. For instance, in \cite{Frolik2013} a 10\% fade depth metric, derived from the cumulative distribution function of the instantaneous received SNR, is proposed to measure severe fading conditions. In \cite{Rao2015, Romero2017b} other metrics associated to the average capacity loss with respect to the Rayleigh fading are discussed. 


The contribution of this work is three-fold: first, in order to analyze the fading severity and using the Rayleigh model as a benchmark, the hyper-Rayleigh character of a fading model is quantified by means of three metrics: amount of fading, asymptotic outage probability and asymptotic average capacity. 
Using these metrics, a four-level scale is proposed to quantify the fading severity of a fading model: full/strong/weak/no hyper-Rayleigh behavior. Second, we provide new analytical results for some statistics of the FTR fading model. Specifically, we obtain closed-form expressions for the $k^{th}$ moment of the SNR and for the amount of fading, as well as asymptotic approximations to the CDF and the average capacity. Third, using this new set of results, we investigate the interplay between fading severity and system performance under FTR fading and we show that the FTR fading model exhibits hyper-Rayleigh behavior in different circumstances, and that it always yields a worse system performance than its TWDP counterpart.


The remainder of this paper is organized as follows. In Section II, we introduce the notation and give some definitions related to fading model metrics. In Section III, additional definitions now focused on the characterization of the hyper-Rayleigh behavior are provided. Section IV is devoted to the analysis of the FTR model, including mathematical derivations to obtain the new results (whose proofs are outlined in the appendices), and the exploration of the interplay between the model parameters and the fading severity by determining its hyper-Rayleigh regime. In Section V, numerical results are presented in which several settings of the FTR model are discussed. Finally, conclusions are drawn in Section VI.

\section{Definitions}
\label{S2}
Throughout this paper, $\mathbb{E}\{\cdot\}$ denotes the expectation operator. The symbol $\sim$ reads as \emph{statistically distributed as}. The instantaneous signal-to-noise ratio (SNR) $\gamma$ is related to the received signal envelope $r$ as  $\gamma=\overline\gamma\frac{|r|^2}{\mathbb E\{|r|^2\}}$, where $\overline\gamma=\mathbb{E}[\gamma]$ is the average SNR. Uniformly distributed random variables (RVs) in the interval $[a,b)$ are denoted as $z\sim\mathcal{U}[a,b)$. The probability density function (PDF) of the RV $\gamma$ is denoted as $f_{\gamma}(\cdot)$, whereas the cumulative distribution function (CDF) of the RV $\gamma$ is denoted as $F_{\gamma}(\cdot)$.
\vspace{1mm}\\



\begin{definition}[Normalized moments of the SNR]\\
The normalized $k^{th}$ moment of $\gamma$ is defined as
\begin{equation}
\label{mom}
\mathcal{M}(k)\triangleq
\frac{\E\{\gamma^k\}}{\overline\gamma^k}=\int_{0}^{\infty} \left(\frac{\gamma}{\overline\gamma}\right)^k f_{\gamma}(\gamma)d\gamma.
\end{equation}
\end{definition}

\begin{definition}[Amount of Fading]\\
The Amount of Fading (AoF) is defined as \cite{Simon2005}
\begin{equation}
\label{AF}
{\rm{AoF}}=\frac{\mathbb{E}[\gamma^2]}{\overline{\gamma}^2}-1=\mathcal{M}(2)-1,
\end{equation}
for the instantaneous SNR $\gamma$.
\end{definition}

\begin{definition}[Outage Probability]\\
The Outage Probability (OP) is defined as the probability that the instantaneous SNR $\gamma$ falls below a given threshold value $\gamma_{\rm th}$, i.e. $OP=\Pr\{\gamma<\gamma_{\rm th}\}$. Hence, it can be computed by evaluating the CDF of the SNR, as 
\begin{equation}
\label{OP}
{\rm{OP}}=F_{\gamma}(\gamma_{\rm th}).
\end{equation}
\end{definition}

\begin{definition}[Asymptotic Outage Probability]\\
For $\gamma_{\rm th}<<\overline\gamma$, the OP can be approximated as \cite{Wang2003}
\begin{equation}
\label{aOP}
{\rm{OP}} \approx \Delta_{\rm{PO}} \left(\frac{\gamma_{\rm th}}{\overline\gamma}\right)^d\triangleq{\rm{aOP}},
\end{equation}
where $d>0$ is usually referred to as \emph{diversity order}, and $\Delta_{\rm{PO}}>0$ is a power offset factor.
\end{definition}

\begin{definition}[Average Capacity]\\
The average capacity $\overline{C}$ per unit bandwidth is defined as the instantaneous Shannon capacity averaged over all possible fading states, i.e. 
\begin{equation}
\label{AC}
\overline{C}=\int_{0}^{\infty} \log_2\left(1+\gamma\right)f_{\gamma}(\gamma)d\gamma,\hspace{8mm} ({\rm bps/Hz}).
\end{equation}
\end{definition}

\begin{definition}[Asymptotic Average Capacity]\\
For sufficiently large $\overline\gamma$, the average capacity can be approximated as
\begin{equation}
\label{aAC}
\overline{C}\approx \log_2(\overline\gamma) - L \triangleq {\rm a}\overline{C},
\end{equation}
where $L\geq0$ due to Jensen's inequality, and can be computed from \eqref{mom} as in \cite{Yilmaz2012}
\begin{equation}
L\triangleq -\log_2(e)\left.\frac{d\mathcal{M}(k)}{dk}\right\rvert_{k=0} =-\log_2(e)\mathcal{M}'(0).
\end{equation}
Note that $L=0$ in the absence of fading, i.e. the additive white Gaussian noise (AWGN) case.
\end{definition}

%
%

\section{Quantifying the hyper-Rayleigh Regime}
\label{S3}

In this Section, we give a formal definition of the hyper-Rayleigh regime in the context of fading channels. Specifically, we define a set of performance metrics related to the actual fading channel distribution, compared to that of the Rayleigh case. These metrics, which are introduced next, are based on the amount of fading, outage probability and average capacity.

\begin{definition}[Hyper-Rayleigh regime in the AoF sense]\\
Let $\gamma$ be the instantaneous SNR under an arbitrary fading model $\mathcal{X}$. Then, we say that the distribution $\mathcal{X}$ has hyper-Rayleigh behavior in the AoF sense if the following condition holds for a certain set of parameter values:
\begin{equation}
\label{hRAoF}
{\rm{AoF}}^{\mathcal{X}}>{\rm{AoF}}^{\rm{Ray}}=1.
\end{equation}
\end{definition}

The previous definition implies that the fading model $\mathcal{X}$ is hyper-Rayleigh if the AoF is larger than its Rayleigh counterpart, for which ${\rm{AoF}}^{\rm{Ray}}=1$, a well-known result \cite{Simon2005}.

\begin{definition}[Hyper-Rayleigh regime in the OP sense]\\
Let $\gamma$ be the instantaneous SNR under an arbitrary fading model $\mathcal{X}$. Then, we say that the distribution $\mathcal{X}$ has hyper-Rayleigh behavior in the OP sense if 
\begin{align}
\label{hRaOP}
&{\rm{aOP}}^{\mathcal{X}}(\gamma_{\rm th})={\rm{aOP}}^{\rm{Ray}}(\gamma_{\rm th}) \implies \overline{\gamma}^{\mathcal{X}}> \overline{\gamma}^{\rm{Ray}},
\end{align}
for a certain set of parameter values.
\end{definition}

\begin{figure}[!t]
\centering
\includegraphics[width=0.97\columnwidth]{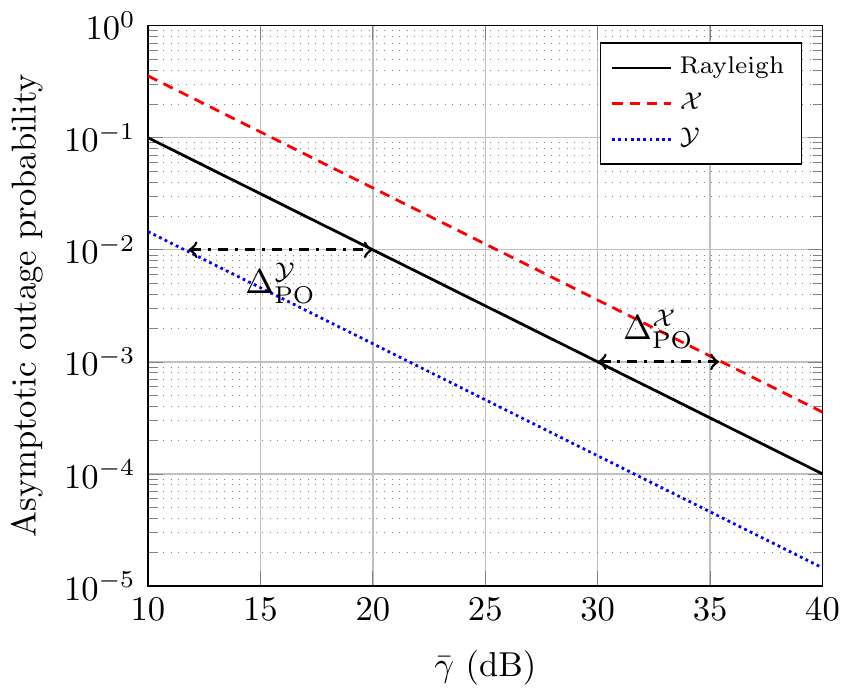}
\caption{Calculation of the power offset metric $\Delta_{\rm PO}({\rm dB})$ from the asymptotic OP, in two different situations. The fading channel model $\mathcal{X}$ has hyper-Rayleigh behavior in the OP sense, unlike the fading channel model $\mathcal{Y}$.}
\label{fig:PO}
\end{figure}

Because the asymptotic OP in the Rayleigh case is given by setting $d=1$ and $\Delta_{\rm{PO}}^{\rm{Ray}}=1$ in \eqref{aOP}, we can easily see that the hyper-Rayleigh behavior is governed by the diversity order of the distribution $\mathcal{X}$. In case $d^{\mathcal{X}}>1$, the asymptotic decay of the OP is faster than in the Rayleigh case and the condition in \eqref{hRaOP} is not met for aOP. Similarly, if $d^{\mathcal{X}}<1$ then the OP under $\mathcal{X}$ fading decays slower than the OP under Rayleigh fading, and hence the distribution $\mathcal{X}$ can be regarded as hyper-Rayleigh. Finally, for $d^{\mathcal{X}}=d^{\rm{Ray}}=1$ we can express the condition in \eqref{hRaOP} as
\begin{align}
10 \log_{10} \left(\frac{\gamma_{\rm th}}{\overline\gamma^{\rm Ray}}\right)&= 10 \log_{10} \left(\frac{\gamma_{\rm th}}{\overline\gamma^{\mathcal{X}}}\right) + 10 \log_{10} \left(\Delta_{\rm PO}^{\mathcal{X}}\right),
\end{align}
so that
\begin{equation}
\label{eqPO}
\Delta_{\rm PO}^{\mathcal{X}}({\rm dB})= 10 \log_{10} \left(\frac{\overline\gamma^{\mathcal{X}}}{\overline\gamma^{\rm Ray}}\right) >0.
\end{equation}
We note that the power offset metric in \eqref{eqPO} is similar to the empirical $10\%$ fade depth metric defined in \cite{Frolik2013}. The hyper-Rayleigh behavior in this case appears when $\Delta_{\rm PO}^{\mathcal{X}}({\rm dB})>0$. The interpretation of \eqref{eqPO} in this situation is exemplified in Fig. \ref{fig:PO}. We see that for a target OP, the power offset metric for the fading channel $\mathcal{X}$ is roughly $\Delta_{\rm PO}^{\mathcal{X}}({\rm dB})\approx5.1$dB; this implies that in order to reach the same OP as in the Rayleigh case, 5.1dB more are required for the average SNR under $\mathcal{X}$-fading. Conversely, for the case of considering the fading model $\mathcal{Y}$, we see that $\Delta_{\rm PO}^{\mathcal{Y}}({\rm dB})\approx-8.4$dB, i.e. we can have the same OP as in the Rayleigh case with a smaller average SNR. Note that because the power offset metric is defined from the asymptotic OP, it does not depend on the target OP value used in \eqref{hRaOP}.

\begin{definition}[Hyper-Rayleigh regime in the average capacity sense]\\
Let $\gamma$ be the instantaneous SNR under an arbitrary fading model $\mathcal{X}$. Then, we say that the distribution $\mathcal{X}$ has hyper-Rayleigh behavior in the average capacity sense if the following condition holds for a certain set of parameter values.
\begin{equation}
\label{hRaC}
\overline{\gamma}^{\mathcal{X}}= \overline{\gamma}^{\rm{Ray}} \implies {\rm a}\overline{C}^{\mathcal{X}}<{\rm a}\overline{C}^{\rm Ray}
\end{equation}
\end{definition}

\begin{figure}[!t]
\centering
\includegraphics[width=0.97\columnwidth]{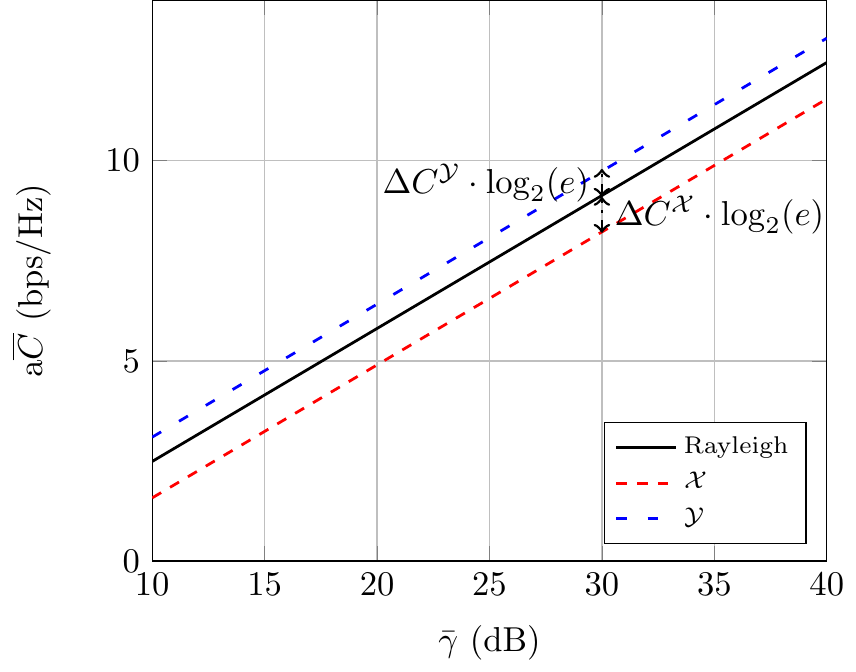}
\caption{Calculation of the capacity offset metric $\Delta{C}$ from the asymptotic capacity, in two different situations. The fading channel model $\mathcal{X}$ has hyper-Rayleigh behavior in the average capacity sense, unlike the fading channel model $\mathcal{Y}$.}\label{fig:C}
\end{figure}

The asymptotic average capacity under Rayleigh fading is obtained \cite{Yilmaz2012} by setting $L^{\rm Ray}=\log_2(e)\cdot\gamma_e$ in \eqref{aAC}, where $\gamma_e=0.577215...$ is the Euler-Mascheroni constant. Hence, the condition in \eqref{hRaC} can be expressed from \eqref{aAC} as
\begin{align}
\label{hRaC2}
&L^{\mathcal{X}}>L^{\rm Ray}\implies\Delta{C}^{\mathcal{X}}\triangleq -\gamma_e-\left.\frac{dM^{\mathcal{X}}(k)}{dk}\right\rvert_{k=0}>0,
\end{align}
which is directly computed by evaluating the first derivative of the $k^{th}$ moment of the fading distribution with respect to $k$, evaluated at $k=0$. The interpretation of \eqref{hRaC2} is exemplified in Fig. \ref{fig:C}. Let us first consider the fading model $\mathcal{X}$; we see that for a given average SNR, the capacity in the high SNR regime is lower than in the Rayleigh case. Hence, the capacity loss metric $\Delta{C}^{\mathcal{X}}\approx0.63>0$, which indicates an hyper-Rayleigh behavior in the average capacity sense. Conversely, the capacity loss metric $\Delta{C}^{\mathcal{Y}}\approx-0.42<0$, and hence in this case we do not have hyper-Rayleigh behavior.

This set of metrics allows to characterize the hyper-Rayleigh behavior in a complete form. As we will show in the next Section, the three conditions may not be met at the time for a given distribution. For this reason, we propose the following intensity scale to measure the \emph{hyper-Rayleighness} of a fading channel: we will say that a fading channel has \emph{full hyper-Rayleigh} behavior if it meets all the three conditions defined in Propositions 1-3, i.e., it is hyper-Rayleigh in the AoF, OP and average capacity senses. We will say that a fading channel has \emph{strong hyper-Rayleigh} behavior if it meets two of the three conditions defined in Definitions 7-9. Finally, we will say that a fading channel has \emph{weak hyper-Rayleigh} behavior if it meets only one of the three conditions defined in Definitions 7-9. In case none of the conditions is met, then the fading model does not exhibit hyper-Rayleigh behavior.

\section{A Case Study: FTR Fading}
In this Section, we will use the previously defined set of metrics to determine to what extent the recently proposed FTR fading model has hyper-Rayleigh behavior. We will also pay attention to some of the special cases of the FTR fading model, which will provide important insights on the hyper-Rayleigh behavior of very relevant fading models in the literature.

\label{S4}
\subsection{System Model for FTR fading}
The received signal in a wireless scenario can be expressed in the following general form \cite{Durgin2002}:
\begin{equation}
\label{eqGeneral}
r  = \sum_{i=1}^N V_i  \exp \left( {j\varphi _i } \right) + \sum_{i=1}^M A_i  \exp \left( {j\phi _i } \right),
\end{equation}
where the $V_i$ and $A_i$ indicate constant amplitudes for each of the multipath waves, which have random phases $\{\varphi_i,\phi_i\}\sim\mathcal{U}[0,2\pi)$. For a sufficiently large number of waves of relatively low amplitudes, and assuming that such waves are included in the second term in \eqref{eqGeneral}, the central limit theorem applies as $M\rightarrow\infty$ and therefore it can be approximated by a complex Gaussian RV $V_{d}=X+jY$ with $\mathbb E\{|V_{d}|^2\}=2\sigma^2$, which will be regarded as the \emph{diffuse component}. Under this premise, we can rewrite \eqref{eqGeneral} as
\begin{equation}
\label{eqGeneral2}
r  = \sum_{i=1}^N V_i  \exp \left( {j\varphi _i } \right) + V_d.
\end{equation}
The $N$ remaining waves correspond to a set of dominant or specular waves usually regarded to as \emph{LoS} components; this will be the convention throughout the rest of the paper. For $N=2$, the popular and versatile TWDP fading model emerges; hence, the instantaneous SNR $\gamma$ under TWDP fading is fully determined by its average SNR $\overline\gamma$, and the parameters $K = \frac{{V_1^2  + V_2^2 }}{2\sigma^2}$ and $\Delta  = \frac{{2V_1 V_2 }}
{{V_1^2  + V_2^2 }}$; i.e. we say that $\gamma\sim\mathcal{T}(\overline{\gamma}; K, \Delta)$.

The FTR fading model in \cite{Romero2017} arises as a natural generalization of the TWDP, on which the LoS components are allowed to randomly fluctuate. Hence, the received signal under FTR fading can be expressed as:
\begin{equation}
\label{eq:04}
r  = \sqrt \zeta V_1  \exp \left( {j\varphi _1 } \right) + \sqrt \zeta V_2  \exp \left( {j\varphi _2 } \right) + V_{d},
\end{equation}
where $V_1$ and $V_2$ are the constant amplitudes of the dominant specular waves with random phases, $V_{d}$ denotes the diffuse component and $\zeta$ is in charge of modeling the amplitude fluctuations in the LoS components, and is assumed to follow a Gamma distribution with unit mean and shape parameter $m>0$, i.e. $\zeta\sim \mathcal{G}(m,1/m)$ with PDF given by
\begin{equation}
\label{PDF_Gamma}
f_{\zeta}(u)=\frac{m^{m}}{\Gamma(m)}u^{m-1}e^{-m u}.
\end{equation}

When conditioned to $\zeta$ the instantaneous SNR under FTR fading is distributed according to the TWDP distribution. This observation allowed to fully characterize the PDF and CDF under FTR fading with arbitrary real $m$ by either an inverse Laplace transform over the MGF \cite{Romero2017} or in the form of an infinite mixture of gamma PDFs/CDFs \cite{Zhang2018}. Additional simplified forms were available for the specific case of $m$ being an integer, as well as simple closed-form approximations \cite{Romero2017}. In the next subsection, we provide new analytical closed-form results for some statistics of the FTR fading model, which will be of later use to characterize the hyper-Rayleigh behavior of this class of fading channels.

\subsection{New Results for FTR fading}
The new statistics of the FTR distribution derived in this work include an asymptotic approximation for the CDF, the moments of the SNR, the AoF metric and the asymptotic average capacity. These results hold for an arbitrary choice of the fading parameters $K$, $\Delta$ and $m$.

\lemma[{Asymptotic approximation for the CDF}]{Let $\gamma$ be the instantaneous SNR under FTR fading. Then, for $K<\infty$ an asymptotic approximation for the FTR CDF can be given as:}
\begin{equation}
\label{eqLemma0}
F_{\gamma}(\gamma)\approx \frac{\gamma}{\overline\gamma} \underbrace{\frac{(1+K)}{(1+\frac{K}{m})^m}{}_2F_1\left(\frac{m}{2},\frac{1+m}{2};1;\frac{\Delta^2}{(\frac{m}{K}+1)^2}\right)}_{\Delta_{\rm PO}^{\rm FTR}},
\end{equation}
where ${}_2F_1(\cdot)$ is the Gauss hypergeometric function.
\begin{IEEEproof}
See Appendix \ref{app1}.
\end{IEEEproof}

\lemma[Normalized moments]{Let $\gamma$ be the instantaneous SNR under FTR fading. Then, the $k^{th}$ normalized moment of $\gamma$ can be expressed as}
\begin{equation}
\label{eqLemma1}
\mathcal{M}(k)=\frac{k!}{(1+K)^k} \sum_{i=0}^{k}\binom{k}{i}\frac{K^i}{i!} {}_2F_1\left(\tfrac{1}{2},inc-i;1;\tfrac{2\Delta}{\Delta-1}\right)G_{i,m}.
\end{equation}
where $G_{i,m}=m^{-i}\frac{\Gamma(m+i)}{\Gamma(m)}$, and $\Gamma(\cdot)$ is the Gamma function.
\begin{IEEEproof}
See Appendix \ref{app2}.
\end{IEEEproof}

The expression for the moments of the SNR in FTR fading is a new result in the literature to the best of the authors' knowledge. A simplified expression can be obtained for the case of $\Delta=1$, which is given in the following corollary.

\corollary{Let $\gamma$ be the instantaneous SNR under FTR fading. Then, if $\Delta=1$ the $k^{th}$ normalized moment of $\gamma$ can be expressed as}
\begin{equation}
\label{eqCorollary1}
\mathcal{M}(k)=\frac{k!}{(1+K)^k} \sum_{i=0}^{k}\binom{k}{i}\frac{(2K)^i}{\sqrt{\pi}m^ii!} \frac{\Gamma(i+1/2)}{\Gamma(i+1)}\frac{\Gamma(m+i)}{\Gamma(m)}.
\end{equation}
\begin{IEEEproof}
Following the same steps as in the proof of Lemma 1, and noting that
\begin{equation}
\label{eqProofA}
\frac{1}{\pi}\int_{0}^{\pi}\left(1+\Delta\cos\theta\right)^id\theta=\frac{2^i}{\sqrt{\pi}}\frac{\Gamma(i+1/2)}{\Gamma(i+1)},
\end{equation}
the proof is completed.
\end{IEEEproof}

%

As a by-product of these expressions for the moments, a closed-form expression for the AoF can be obtained as follows.

\lemma[Amount of Fading]{Let $\gamma$ be the instantaneous SNR under FTR fading. Then, the AoF in FTR fading is given by
\begin{equation}
\label{AoF}
{\rm{AoF}}=1-\underbrace{\left(\frac{K}{1+K}\right)^2}_{p(K)}\left[2-\underbrace{\left(1+\frac{\Delta^2}{2}\right)}_{q(\Delta)}\underbrace{\left(1+\frac{1}{m}\right)}_{r(m)}\right].
\end{equation}
\begin{IEEEproof}
Using the expression of the moments in \eqref{eqLemma1} for $k=2$, the special values of the second moment in \cite[eq. 8]{Lopez2016}, and noting that $\frac{\Gamma(m+2)}{\Gamma(m)m^2}=1+\frac{1}{m}$, the desired expression is obtained after some manipulations. 
\end{IEEEproof}

Expression \eqref{AoF} is a new result in the literature to the best of our knowledge. In Table \ref{table_1}, the AoF of most relevant special cases included in the FTR fading model is summarized. These include the TWDP (i.e., $m\rightarrow\infty$) and the fluctuating two-wave (FTW) (i.e., $K\rightarrow\infty$) \cite{Rao2015}, the two-wave (i.e., $K\rightarrow\infty$, $m\rightarrow\infty$), the Rician shadowed (i.e., $\Delta=0$), Rician (i.e., $\Delta=0$, $m\rightarrow\infty$), Hoyt (i.e., $q=\sqrt{\frac{1+K(1-\Delta)}{1+K(1+\Delta)}}$ and $m=1$) and Rayleigh (i.e., $\Delta=0$, $K=0$, $\forall m$) fading models. 

\begin{table}[!t]
\renewcommand{\arraystretch}{1.7}
\caption{{\rm{AoF} of the FTR fading model and special cases}}
\label{table_1}
\centering
\begin{tabular}
{c|c}
\hline
\hline
Channel model  & {\rm{AoF}}\\
\hline
\hline
FTR & ${\rm{AoF}}=1-\left(\frac{K}{1+K}\right)^2\left[2-\left(1+\frac{\Delta^2}{2}\right)\left(1+\frac{1}{m}\right)\right]$\\
\hline
TWDP &  ${\rm{AoF}}=1-\left(\frac{K}{1+K}\right)^2\left[1-\frac{\Delta^2}{2}\right]$ \\
\hline
FTW  &  ${\rm{AoF}}= \frac{1}{m} \left(1+\frac{\Delta^2}{2}\right) + \frac{\Delta^2}{2}$ \\
\hline
Two-Wave &   ${\rm{AoF}}=\frac{\Delta^2}{2}$\\
\hline
Rician shadowed &  ${\rm{AoF}}=1-\left(\frac{K}{1+K}\right)^2\left(1-\frac{1}{m}\right)$\\
\hline
Rician &  ${\rm{AoF}}=1-\left(\frac{K}{1+K}\right)^2$ \\
\hline 
Hoyt & ${\rm{AoF}}=\frac{2(1+q^4)}{(1+q^2)^2}$\\
\hline
Rayleigh &  ${\rm{AoF}}=1$ \\
\hline
\end{tabular}
\end{table}

\lemma[Asymptotic Capacity]{Let $\gamma$ be instantaneous SNR under FTR fading. Then, the average capacity in the high-SNR regime is tightly approximated by
\begin{equation}
\label{eqCFTR}
{\rm{a}}\overline{C}=\log_2 \overline\gamma - L^{\rm FTR}(K,\Delta,m),
\end{equation}
where
\begin{equation}
\label{eqCFTR2}
L^{\rm FTR}(K,\Delta,m)=\underbrace{\log_2(e)\gamma_e}_{L^{\rm Ray}}-\frac{\log_2(e)}{\pi}\int_{0}^{\pi}\mathcal{F}(\theta)d\theta,
\end{equation}
and $\mathcal{F}(\theta)$ is given at the top of next page in \eqref{eqCFTR3}, with ${}_3F_2(\cdot)$ being a generalized hypergeometric function \cite[eq. (16.2.1)]{NIST} and $\log$ denoting the natural logarithm.
\begin{figure*}[t]
\begin{equation}
\label{eqCFTR3}
\mathcal{F}(\theta)=\log\left(\frac{K(1+\Delta\cos\theta)+m}{m(1+K)}\right)+\frac{K(1+\Delta\cos\theta)(m-1)}{K(1+\Delta\cos\theta)+m}{}_3F_2\left(1,1,2-m;2,2;\frac{K(1+\Delta\cos\theta)}{K(1+\Delta\cos\theta)+m}\right).
\end{equation}
\hrulefill
\end{figure*}
\begin{IEEEproof}
See Appendix \ref{app3}.
\end{IEEEproof}

\subsection{Exploring the hyper-Rayleigh Regime of FTR Fading}

We now use the previously derived analytical results to investigate the effect of the FTR fading parameters, namely $K$, $\Delta$ and $m$, on the fading severity, i.e. to determine its hyper-Rayleigh behavior. 

Let us first begin with the AoF metric. Interestingly, the AoF dependence on the fading model parameters is encapsulated in the functions $p(K)$, $q(\Delta)$ and $r(m)$ as indicated in \eqref{AoF}. Since all these three ancillary functions are positive, this facilitates to understand the interplay between the fading model parameters and the AoF. Specifically, evaluating the first derivative of the AoF with respect to each of the fading parameters allows us to determine the monotonic behavior (increasing or decreasing) of the AoF. These derivatives can be easily computed as follows:
\begin{align}
\frac{\partial {\rm AoF}}{\partial K}&=-\frac{2K}{(1+K)^3}\left(2-q(\Delta)r(m)\right),\\
\frac{\partial {\rm AoF}}{\partial \Delta}&=p(K)r(m)\Delta,\\
\frac{\partial {\rm AoF}}{\partial m}&=-p(K)q(\Delta)\frac{1}{m^2},
\end{align}
We see that $\frac{\partial {\rm AoF}}{\partial \Delta}>0$ $\forall \Delta$; hence, this implies that the AoF is always increased with $\Delta$, i.e. it is always larger than that for $\Delta=0$. Similarly, we also see that $\frac{\partial {\rm AoF}}{\partial m}<0$ $\forall m$. Hence, the AoF is always increased as $m$ is reduced, i.e. as the LoS fluctuation is heavier. This implies that the FTR fading model always has a larger AoF than its TWDP counterpart. Finally, in order to understand the impact of increasing $K$, two different cases need to be studied depending on whether $q(\Delta)r(m)\lessgtr 2$. After some algebra, we see that $\frac{\partial {\rm AoF}}{\partial K}>0$ if $m<\frac{1+\Delta^2/2}{1-\Delta^2/2}$. This means that the AoF increases with $K$ for those values of $m$ lower than the previous threshold value, which depends on $\Delta$. For instance, the AoF increases with $K$ if $m<1$ and ${\Delta=0}$, or with $m<3$ if $\Delta=1$. In these situations, increasing the LoS power is detrimental for the AoF. As it will become clear in Section \ref{S5}, it is evident that the condition in \eqref{hRAoF} is met for some values of the FTR fading model parameters. Hence, we can anticipate that the FTR fading model has hyper-Rayleigh behavior in the AoF sense. However, we also observe that in the TWDP case (i.e. $m\rightarrow\infty$), the AoF is always reduced as $K$ is increased. Direct inspection of the AoF metric for the TWDP case reveals that $\rm AoF^{TWDP}\leq1$ in all instances. Hence, we see that the TWDP fading model is not hyper-Rayleigh in the AoF sense, an observation that has not been made in the literature to the best of our knowledge.


We will now move to the analysis of the hyper-Rayleigh behavior in the OP sense. Because the Two-Wave fading model ($K\rightarrow\infty$, $m\rightarrow\infty$) is a special case of the FTR fading model, and its diversity order is $d=1/2$ for $\Delta=1$ \cite{Eggers2019}, we know beforehand that the FTR fading model also exhibits a hyper-Rayleigh behavior in the OP sense. In the general case, the power offset metric in \eqref{hRaOP} is given by
\begin{equation}
\label{eqPO2}
\Delta_{\rm PO}^{\rm{FTR}}({\rm dB})= 10 \log_{10} \left(\tfrac{(1+K)}{(1+\tfrac{K}{m})^m}{}_2F_1\left(\tfrac{m}{2},\tfrac{1+m}{2};1;\tfrac{\Delta^2}{(\frac{m}{K}+1)^2}\right)\right).
\end{equation}
Although a closed-form expression for this metric is obtained, it is not possible to analytically determine the range of values of $K$, $\Delta$ and $m$ for which the hyper-Rayleigh behavior appears. Hence, we will proceed to the numerical evaluation of this metric in Section \ref{S5}, although several special cases will be later  discussed. For the TWDP case, we have
\begin{equation}
\label{eqPO3}
\Delta_{\rm PO}^{\rm{TWDP}}({\rm dB})= 10 \log_{10} \left( (1+K) e^{-K}I_0(K\Delta),\right),
\end{equation}
by direct inspection of \eqref{eqpepe} in the Appendix \ref{app1}.

Finally, we will consider the average capacity metric defined in \eqref{hRaC}. Using \eqref{eqCFTR2} and \eqref{hRaC2}, we can express
\begin{align}
\label{hRaC2b}
\Delta{C}^{\rm FTR}=-\frac{1}{\pi}\int_{0}^{\pi}\mathcal{F}(\theta)d\theta,
\end{align}
so that all the potential hyper-Rayleigh behavior is encapsulated in one single term, which can be easily evaluated numerically. For the TWDP case, the capacity loss metric reduces to the expression given in \cite{Rao2015}. As stated in \cite{Rao2015}, the average capacity under TWDP fading may be lower than that of the Rayleigh fading channel. Hence, and because the TWDP fading channel is a special case of the FTR model, then the FTR fading model will exhibit hyper-Rayleigh behavior in the average capacity sense. Thus, we can conclude that the FTR fading model has \emph{full} hyper-Rayleigh behavior, whereas the TWDP fading model has \emph{strong} hyper-Rayleigh behavior.

\section{Numerical Results}
\label{S5}

\begin{figure}[!t]
\centering
\includegraphics[width=0.97\columnwidth]{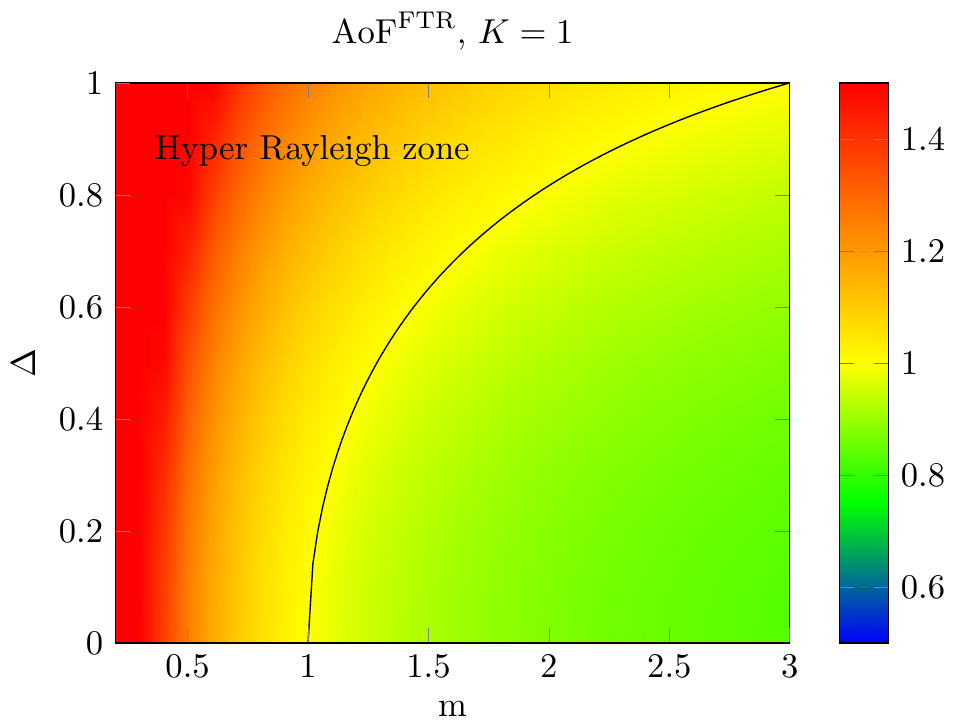}
\caption{Evolution of the AoF for the FTR model, for low LoS (i.e. $K=1$). Solid black line indicates the boundary of the hyper-Rayleigh zone.}
\label{fig:AoF1}
\end{figure}

\begin{figure}[!t]
\centering
\includegraphics[width=0.97\columnwidth]{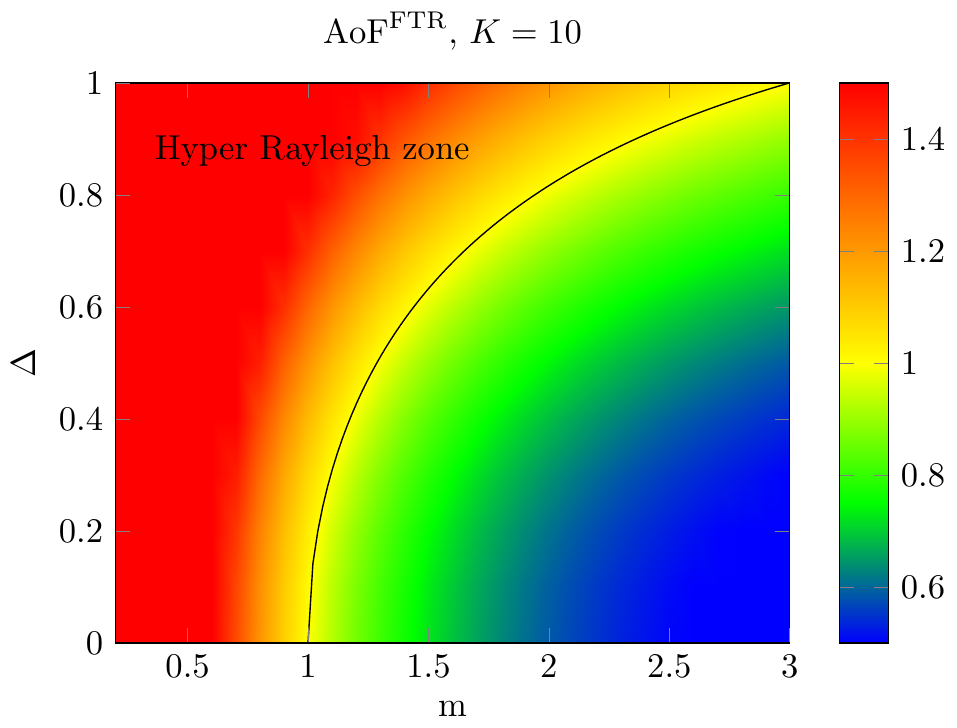}
\caption{Evolution of the AoF for the FTR model, for high LoS (i.e. $K=10$). Solid black line indicates the boundary of the hyper-Rayleigh zone.}
\label{fig:AoF2}
\end{figure}

\begin{figure}[!t]
\centering
\includegraphics[width=0.97\columnwidth]{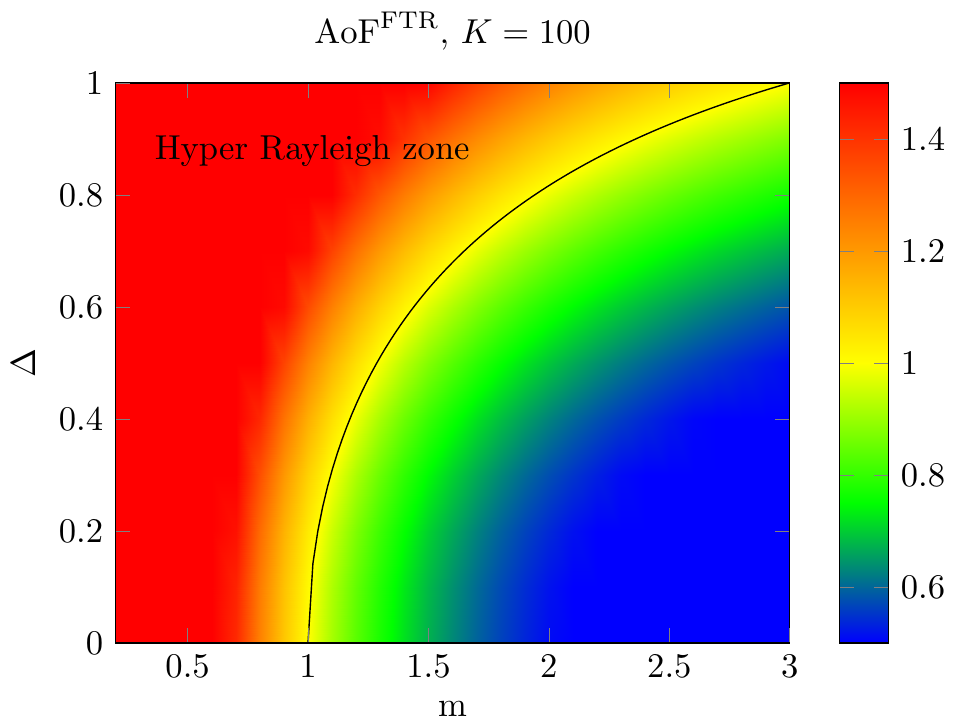}
\caption{Evolution of the AoF for the FTR model, for very high LoS (i.e. $K=100$). Solid black line indicates the boundary of the hyper-Rayleigh zone.}
\label{fig:AoF3}
\end{figure}

We now exemplify the previous approach by providing numerical evidences on the hyper-Rayleigh behavior of the FTR fading model. All results in this section have been double-checked with Monte Carlo simulations, although they are not explicitly superimposed to the figures for the sake of clarity. In the following, we represent the set of metrics defined in Section \ref{S3} as a function of the FTR fading model parameters. For convenience of discussion, we use a 2-D representation using the parameters $m$ and $\Delta$ in the abscissa and ordinate axes, and then change the parameter $K$ accordingly in order to consider three different situations: low LoS (i.e. $K=1$), high LoS (i.e. $K=10$) and very high LoS (i.e. $K=100$). Note that the latter implies that the amount of power due to multipath components is extremely low.

Figs. \ref{fig:AoF1} to \ref{fig:AoF3} represent the AoF of the FTR model given by (\ref{AoF}). We can observe that this model has a clear hyper-Rayleigh behavior in the AoF sense when \mbox{$m<1$}, for any $\Delta$ and $K$. For $1<m<3$, this behavior depends on $\Delta$: increasing the similarity of the specular component powers as (i.e., as $\Delta$ grows) allows to be in the hyper-Rayleigh zone with a lower LoS fluctuation (i.e., a larger $m$). For $m>3$, as predicted by the analytical expressions, there is no hyper-Rayleigh behavior. We see that the parameter $K$ has no influence on the limits of this region, but raising the LoS power implies a higher AoF in the hyper-Rayleigh zone and a lower AoF out of it, as stated in Section \ref{S4}. In the hyper-Rayleigh zone, increasing the LoS power through $K$ is detrimental because it also increases the AoF.

We will now analyze the hyper-Rayleigh behavior in the OP sense. Figs. \ref{fig:aOP1} to \ref{fig:aOP3} show the power offset metric (in dB) related to the asymptotic OP given by (\ref{eqPO2}). Whenever $m<1$, and regardless of $K$ and $\Delta$, the model is in hyper-Rayleigh zone in the OP sense. For $m>1$, this behavior depends on $\Delta$ and $K$. As the power of the two specular components becomes more similar ($\Delta \rightarrow 1$), the fluctuation on these components could be lower (larger $m$) while still being hyper-Rayleigh. As the power in the LoS paths increases (higher $K$), the hyper-Rayleigh behavior still appears when the two specular components are nearly equivalent (i.e. $\Delta$ close to 1), even for lower fluctuations on the LoS components. In fact, we see that the solid black line denoting the boundary of the hyper-Rayleigh region becomes more abrupt as $K$ grows, and the set of parameter values that corresponds to the hyper-Rayleigh region is reduced. It is important to note that increasing $K$ has a notorious influence on the power offset value: for $K=1$, the highest power offset is about 2dB, i.e. in order to reach the same OP as in the Rayleigh case, nearly 2 dB more are required for the average SNR under FTR fading. But for $K=100$, this value could be around 20 dB, a clearly worse situation. 

\begin{figure}[!t]
\centering
\includegraphics[width=0.97\columnwidth]{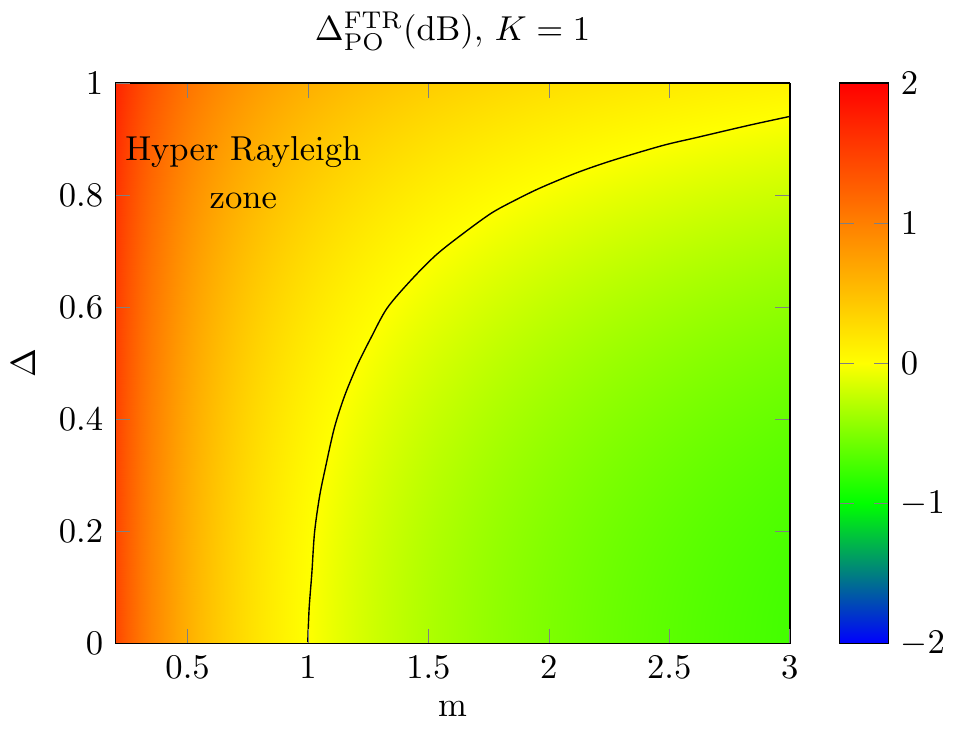}
\caption{Evolution of the power offset metric $\Delta_{\rm PO}^{\rm{FTR}}$(dB) for the FTR model, for low LoS (i.e. $K=1$). Solid black line indicates the boundary of the hyper-Rayleigh zone.}
\label{fig:aOP1}
\end{figure}

\begin{figure}[!t]
\centering
\includegraphics[width=0.97\columnwidth]{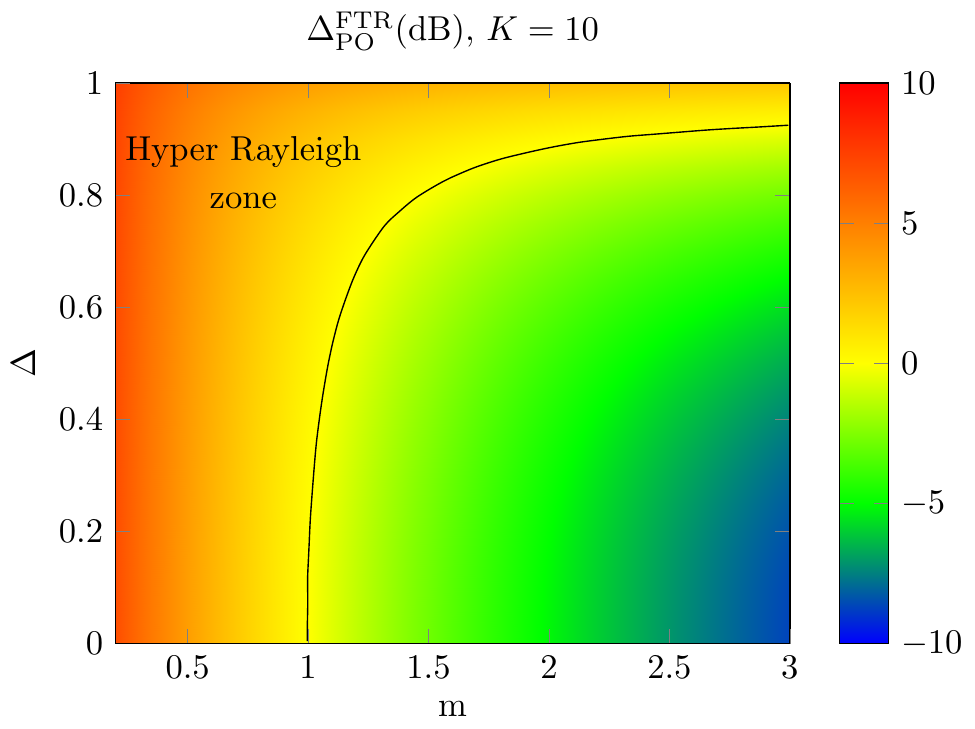}
\caption{Evolution of the power offset metric $\Delta_{\rm PO}^{\rm{FTR}}$(dB) for the FTR model, for high LoS (i.e. $K=10$). Solid black line indicates the boundary of the hyper-Rayleigh zone.}
\label{fig:aOP2}
\end{figure}

\begin{figure}[!t]
\centering
\includegraphics[width=0.97\columnwidth]{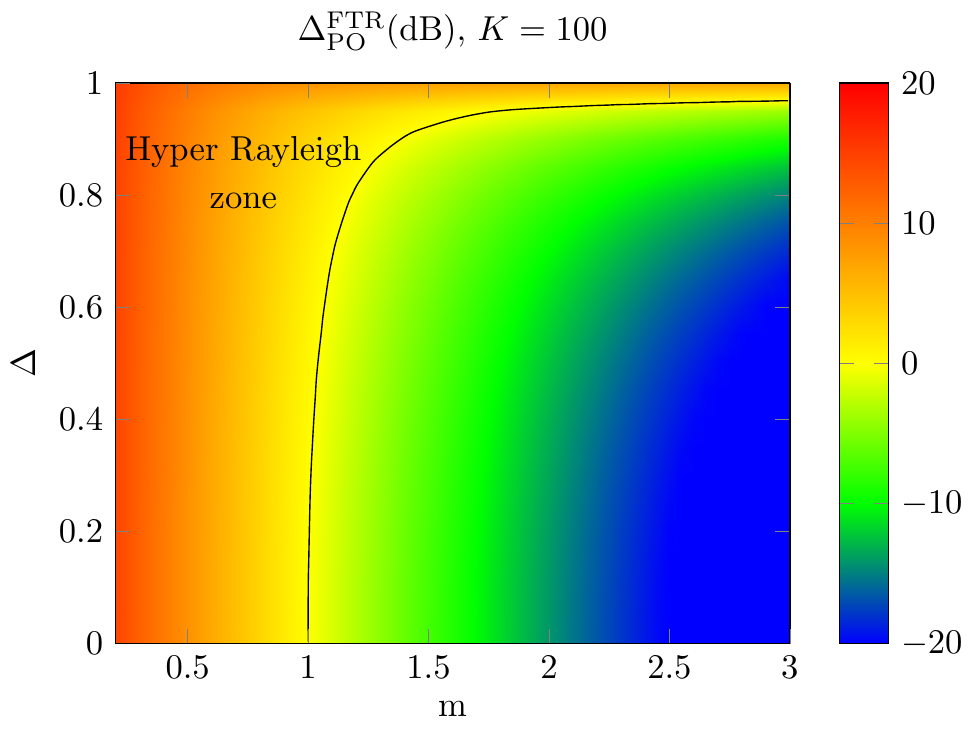}
\caption{Evolution of the power offset metric $\Delta_{\rm PO}^{\rm{FTR}}$(dB) for the FTR model, for very high LoS (i.e. $K=100$). Solid black line indicates the boundary of the hyper-Rayleigh zone.}
\label{fig:aOP3}
\end{figure}

Finally, Figs. \ref{fig:aC1} to \ref{fig:aC3} represent the capacity loss metric given by (\ref{hRaC2}) for the case of the FTR model. As with the other metrics, whenever $m<1$ there is a hyper-Rayleigh behavior, now in the average capacity sense. For other values of $m$, the boundary line changes with $\Delta$ and $K$ similarly to what happens with the power offset metric previously analyzed. We see that the parameter $K$ has a major influence on the maximum capacity loss value. 
This entails that increasing the LoS power in the hyper-Rayleigh zone is detrimental for the capacity, just the opposite behavior experienced in any situation out of the hyper-Rayleigh region. Interestingly, increasing the power of the dominant components helps to be in hyper-Rayleigh region for a slightly lower $\Delta$ than in the low LoS case, and also with a lower LoS fluctuation. In this case, the sets of fading parameters that corresponds to the hyper-Rayleigh region is enlarged. This conforms an area where the AoF in FTR fading is lower than the Rayleigh case, but the capacity of the former is also smaller than under Rayleigh fading.  

We can conclude from the previous study that the FTR fading model has \emph{full} hyper-Rayleigh behavior according to the three performance metrics considered to measure the \emph{hyper-Rayleighness} of a fading channel stated in Section \ref{S3}. This can be summarized in a visually relevant form as in Fig. \ref{figFinal}, on which the three boundaries delimiting the hyper-Rayleigh region for each metric are superimposed for the specific case of $K=100$. We call this representation a \emph{hyper-Rayleigh} map, and encapsulates the hyperRayleigh behavior of the fading model in a unified fashion.


As for some of the special cases of the FTR model, we can make the following observations:
\begin{itemize}
\item TWDP ($m \rightarrow \infty$). As justified in Section \ref{S4}, the TWDP fading model only has \emph{strong} hyper-Rayleigh behavior, since it does not meet the hyper-Rayleigh criterion in the AoF sense.
\item FTW ($K \rightarrow \infty$). Similarly to the regular FTR fading model, this special case also has \emph{full} hyper-Rayleigh behavior because of the LoS fluctuation captured by the parameter $m$. 
\item Two Wave ($K \rightarrow \infty$ and $m \rightarrow \infty$). Even though this model is widely regarded as one of the examples of hyper-Rayleigh behavior \cite{Frolik2007,Rao2015}, using the definitions here proposed we see that it only has \emph{strong} hyper-Rayleigh behavior, and not a \emph{full} one, because of the AoF.
\item Rician shadowed ($\Delta =0$). As we can see from the figures, it can be easily checked that it has \emph{full} hyper-Rayleigh behavior for $m<1$.
\item Rician ($\Delta =0$ and $m \rightarrow \infty$). Confirming the well-known result that the Rician fading is always less detrimental than the case of Rayleigh fading, we see that it has no hyper-Rayleigh behavior.
\item Hoyt ($m=1$). Interestingly, the formulation of the Hoyt model as a special case of the FTR fading model for $m=1$ is useful to confirm that, $\forall q<1$ (i.e. regardless of $K$ and $\Delta$), it has \emph{full} hyper-Rayleigh behavior,
\end{itemize}

\begin{figure}[!t]
\centering
\includegraphics[width=0.97\columnwidth]{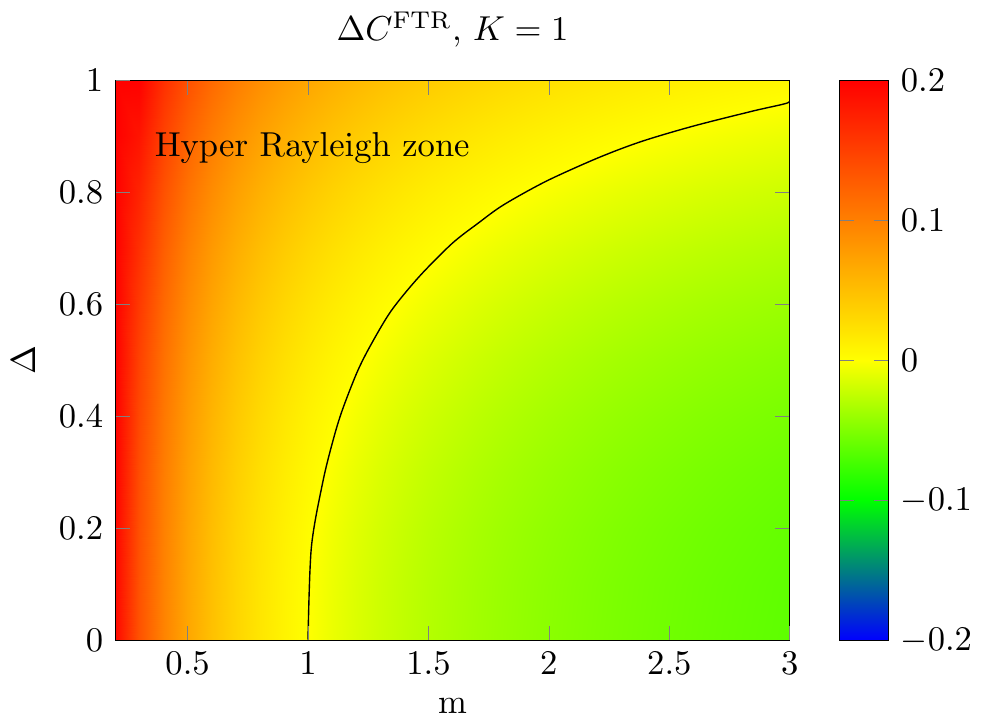}
\caption{Evolution of the capacity offset metric $\Delta C^{\rm{FTR}}$ for the FTR model, for low LoS (i.e. $K=1$). Solid black line indicates the boundary of the hyper-Rayleigh zone.}
\label{fig:aC1}
\end{figure}

\begin{figure}[!t]
\centering
\includegraphics[width=0.97\columnwidth]{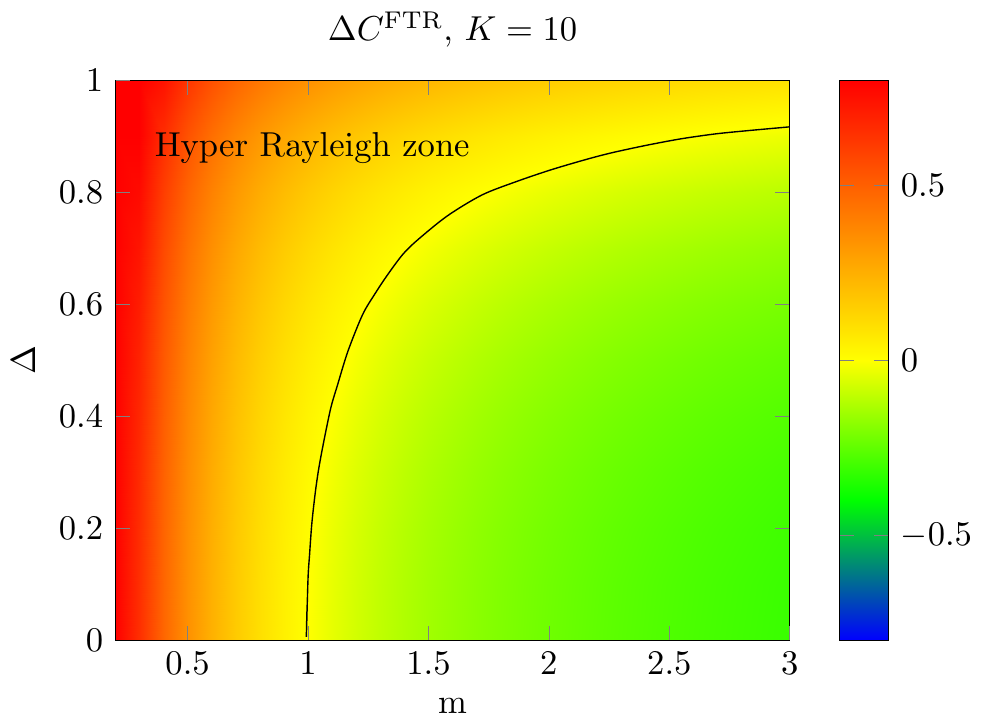}
\caption{Evolution of the capacity offset metric $\Delta C^{\rm{FTR}}$ for the FTR model, for high LoS (i.e. $K=10$). Solid black line indicates the boundary of the hyper-Rayleigh zone.}
\label{fig:aC2}
\end{figure}

\begin{figure}[!t]
\centering
\includegraphics[width=0.97\columnwidth]{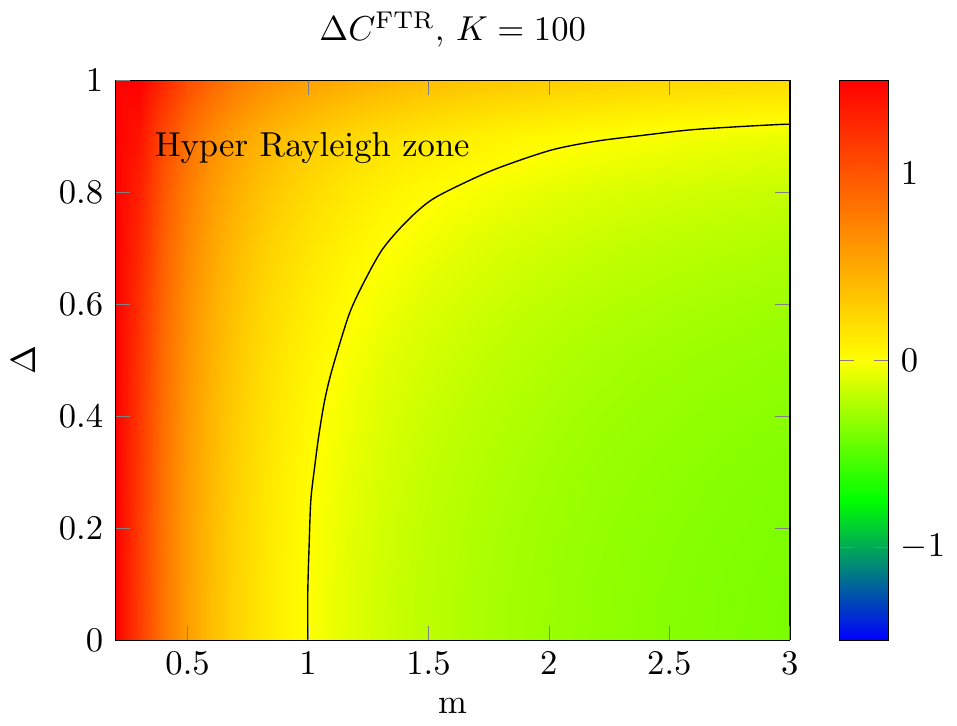}
\caption{Evolution of the capacity offset metric $\Delta C^{\rm{FTR}}$ for the FTR model, for very high LoS (i.e. $K=100$). Solid black line indicates the boundary of the hyper-Rayleigh zone.}
\label{fig:aC3}
\end{figure}

\begin{figure}[!t]
\centering
\includegraphics[width=0.89\columnwidth]{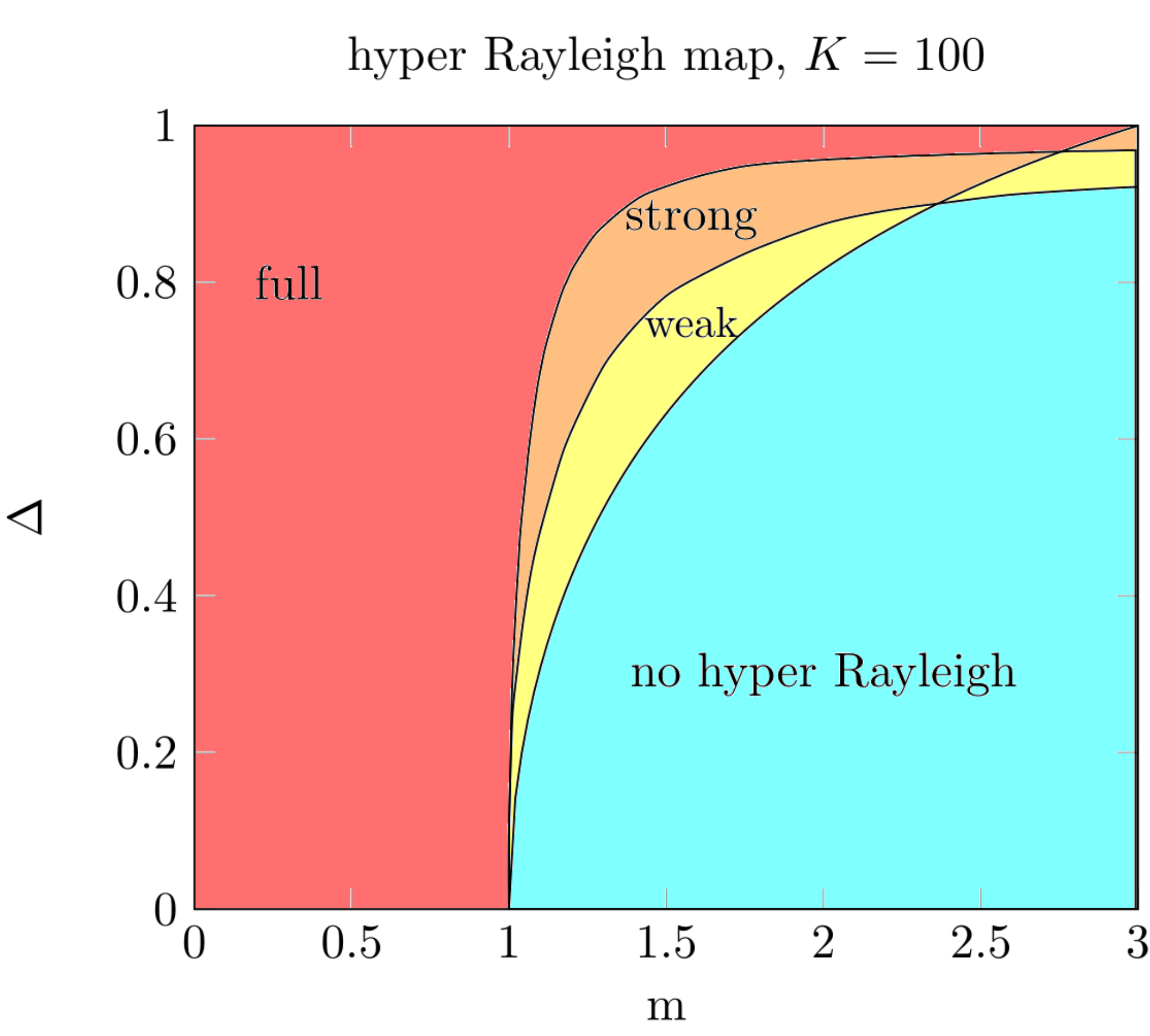}
\caption{Hyper-Rayleigh map for the FTR fading parameters $m$ and $\Delta$, for very high LoS (i.e. $K=100$).}
\label{figFinal}
\end{figure}

\section{Conclusion}
\label{S6}

We have provided a formal definition for a set of conditions that define whether a given fading channel has hyper-Rayleigh or worse-than-Rayleigh behavior. Using this newly proposed definition, we consider the very general FTR fading model to exemplify our approach. We see that some of the fading models which have been classically considered as hyper-Rayleigh, i.e. TWDP and Two-Wave models, do not have \emph{full} hyper-Rayleigh behavior as the AoF metric is never larger than in the Rayleigh case. However, the more general FTR fading model does exhibit \emph{full} hyper-Rayleigh behavior when the dominant specular components heavily fluctuate. A very interesting conclusion extracted is that the random fluctuations in the LoS components captured by this class of fading models allow for modeling more severe fading conditions than its deterministic LoS counterparts.

\begin{appendices}

\section{Proof of Lemma 1}
\label{app1}
An asymptotic approximation for the Rician distribution is given in \cite{Wang2003} as
\begin{equation}
F^{\rm{Rice}}_{\gamma}(\gamma)\approx \frac{\gamma(1+K)}{\overline\gamma} e^{-K}.
\end{equation}
Using the analytical approach in \cite{Rao2015} that connects the TWDP distribution with an underlying Rician distribution of randomly varying parameter $K(\theta)=K(1+\Delta \cos \theta)$, with $\theta\sim\mathcal{U}[0,2\pi)$, we obtain a tail approximation for the TWDP distribution as:
\begin{align}
\label{eqpepe}
F^{\rm{TWDP}}_{\gamma}(\gamma)\approx & \frac{\gamma(1+K)}{\overline\gamma2\pi} \int_{0}^{2\pi}e^{-K(1+\Delta \cos \theta)}d\theta, \nonumber\\
&\approx \frac{\gamma(1+K)}{\overline\gamma} e^{-K}I_0(K\Delta),
\end{align}
where we used the definition of the modified Bessel function of the first kind \cite[32.10.1]{NIST}. 

Let us define the ancillary RV $\gamma_u$ as the instantaneous SNR under FTR fading, conditioned to a given value of $\zeta$, i.e. $\gamma_u\triangleq\gamma|\zeta=u$. Hence, we have that $\gamma_u\sim\mathcal{T}(\overline{\gamma_u}; K_u, \Delta)$, with $\overline\gamma_u=\E\{\gamma_u\}$, $K_u=uK$, with $K = \frac{{V_1^2  + V_2^2 }}{2\sigma^2}$ and $\Delta  = \frac{{2V_1 V_2 }}
{{V_1^2  + V_2^2 }}$. Noting that the following relation holds \cite{Romero2017},
\begin{equation}
\label{eqRomero}
\frac{1+K_u}{\overline\gamma_u}=\frac{1+K}{\overline\gamma},
\end{equation}
we can obtain the asymptotic approximation of the FTR distribution by integrating over all possible values of $\zeta$, as
\begin{align}
F_{\gamma}(\gamma)\approx & \frac{\gamma(1+K)}{\overline\gamma} \int_{0}^{\infty} e^{-Ku}I_0(Ku\Delta) \frac{m^{m}}{\Gamma(m)}u^{m-1}e^{-m u}du.
\end{align}
Solving the integral in \eqref{eqLemma0} using \cite[3.15.1.2]{Prudnikov4} yields the desired result.

\section{Proof of Lemma 2}
\label{app2}
We define the same ancillary variable as in the proof of Lemma 1. From \cite{Lopez2016}, we have that
\begin{equation}
\label{eqProof1}
\E\{\gamma_u^k\}=\frac{k! \overline\gamma_u^k}{(1+K_u)^k 2\pi} \sum_{i=0}^{k}\binom{k}{i}\frac{K_u^i}{i!}\int_{0}^{2\pi}\left(1+\Delta\cos\theta\right)^id\theta.
\end{equation}
Recalling that the integrand is symmetric with respect to $\pi$, we can rewrite \eqref{eqProof1} as
\begin{equation}
\label{eqProof2}
\E\{\gamma_u^k\}=\frac{k! \overline\gamma^k}{(1+K)^k \pi} \sum_{i=0}^{k}\binom{k}{i}\frac{(uK)^i}{i!}\int_{0}^{\pi}\left(1+\Delta\cos\theta\right)^id\theta.
\end{equation}
The inner integral can be solved after a change of variables ${t=\cos\theta}$ and using the integral form of the ${}_2F_1(\cdot)$ hypergeometric function in \cite[eq. 35.7.5]{NIST}, yielding
\begin{equation}
\label{eqProof3}
\frac{1}{\pi}\int_{0}^{\pi}\left(1+\Delta\cos\theta\right)^id\theta=\left(1-\Delta\right)^i {}_2F_1\left(\tfrac{1}{2},-i;1;\tfrac{2\Delta}{\Delta-1}\right).
\end{equation}
Combining \eqref{eqProof3}, \eqref{eqProof2} and using $\E\{\gamma^k\}=\E\{\E\{\gamma_u^k\}\}$, where the outer expectation operation is performed over all the possible values of $\zeta$ with \eqref{PDF_Gamma}, we have
\begin{align}
\label{eqProof4}
\E\{\gamma^k\}=&\frac{k! \overline\gamma^k}{(1+K)^k} \sum_{i=0}^{k}\binom{k}{i}\frac{K^i}{i!} {}_2F_1\left(\tfrac{1}{2},-i;1;\tfrac{2\Delta}{\Delta-1}\right)\nonumber\\ \times&\int_{0}^{\infty}u^i \frac{m^{m}}{\Gamma(m)}u^{m-1}e^{-m u}du.
\end{align}
The integral in \eqref{eqProof4} can be expressed in closed-form as
\begin{equation}
\label{eqProof5}
\int_{0}^{\infty}u^i \frac{m^{m}}{\Gamma(m)}u^{m-1}e^{-m u}du=m^{-i}\frac{\Gamma(m+i)}{\Gamma(m)}.
\end{equation}
Substituting \eqref{eqProof5} into \eqref{eqProof4} and then normalizing to $\overline\gamma^k$ completes the proof, yielding \eqref{eqLemma1}. 

\section{Proof of Lemma 4}
\label{app3}
Let us use \eqref{eq:04} to express the received signal power as
\begin{equation}
\label{eqapp3-01}
r  = \exp \left( {j\varphi_1} \right) \left(\sqrt \zeta V_1 + \sqrt \zeta V_2  \exp \left( {j\alpha} \right) + V_{d}\exp \left( {-j\varphi_1} \right)\right),
\end{equation}
where $\alpha=\varphi_2-\varphi_1$. Because $V_d$ is a complex circularly symmetric RV, its distribution is not affected by the $\exp \left( {-j\varphi_1} \right)$ term. As stated in \cite{Rao2015}, and because of the modulo $2\pi$ operation, then $\alpha\sim\mathcal{U}[0,2\pi)$. Hence, the distribution of the received signal power can be equivalently computed from
\begin{equation}
|r|^2  = |\sqrt \zeta V_1 + \sqrt \zeta V_2  \exp \left( {j\alpha} \right) + V_{d}|^2.
\end{equation}
Now let us define the ancillary RV $\gamma_{\alpha}$ as the instantaneous SNR under FTR fading, conditioned to a given value of $\alpha$, i.e. $\gamma_{\alpha}\triangleq\gamma|\alpha=\theta$. Then, it holds that $\gamma_{\alpha}$ follows a (squared) Rician shadowed distribution \cite{Abdi2003} with 
mean $\overline{\gamma_{\alpha}}=\mathbb{E}[\gamma_{\alpha}]$ and non-negative real shape parameters $K_{\alpha}=K(1+\Delta\cos(\theta))$ and $m$, i.e, $\gamma\sim\mathcal{RS}(\overline{\gamma_{\alpha}}; K_{\alpha}, m)$. Similarly to \eqref{eqRomero}, the following relation holds:
\begin{equation}
\label{eqRomero2}
\frac{1+K_\alpha}{\overline\gamma_\alpha}=\frac{1+K}{\overline\gamma}.
\end{equation}
This allows to express the conditional asymptotic capacity for $\overline\gamma_{\alpha}$ as
\begin{equation}
\label{eqapp3-02}
\overline{C}(\overline{\gamma}_{\alpha})\approx \log_2{\overline\gamma_{\alpha}}-L^{\rm RS}(K_{\alpha},m),
\end{equation}
where $L^{\rm RS}$ is given in \cite[Table II]{Moreno2016} as
\begin{align}
\label{eqapp3-03}
L^{\rm RS}=&{\log_2(e)\gamma_e}-\log_2(e)\left[\log\left(\tfrac{K_{\alpha}+m}{m(1+K_{\alpha})}\right)+\tfrac{K_{\alpha}(m-1)}{K_{\alpha}+m}\right.\nonumber\\ &\left.\times{}_3F_2\left(1,1,2-m;2,2;\frac{K_{\alpha}}{K_{\alpha}+m}\right)\right].
\end{align}
Now plugging \eqref{eqRomero2} into \eqref{eqapp3-02} and \eqref{eqapp3-03}, we can express
\begin{equation}
\label{eqapp3-04}
\overline{C}(\overline{\gamma})\approx \log_2{\overline\gamma}-L_{\alpha}^{\rm RS}(K_{\alpha},m),
\end{equation}
where now the dependence on $\alpha$ of the terms $\overline\gamma_{\alpha}$ and $(1+K_{\alpha})$ is dropped because of \eqref{eqRomero2}, i.e.,
\begin{align}
\label{eqapp3-05}
L_{\alpha}^{\rm RS}=&{\log_2(e)\gamma_e}-\log_2(e)\left[\log\left(\tfrac{K_{\alpha}+m}{m(1+K)}\right)+\tfrac{K_{\alpha}(m-1)}{K_{\alpha}+m}\right.\nonumber\\ &\left.\times{}_3F_2\left(1,1,2-m;2,2;\frac{K_{\alpha}}{K_{\alpha}+m}\right)\right].
\end{align}
Finally, averaging \eqref{eqapp3-04} over all possible values of the RV $\alpha$ yields the desired result for the asymptotic capacity over the FTR fading channel.

\end{appendices}

\bibliographystyle{ieeetr}
\bibliography{FTR6}

\end{document}